%% file: fading-paper.tex
\documentclass[12pt]{article}
\usepackage{amsmath}

\usepackage{epsfig}
\include{preamble}

\begin{document}

\pagenumbering{arabic}
\renewcommand{\textfraction}{0}
\title{Writing on Fading Paper\thanks{These results were first mentioned briefly in Allerton Conference, October 2004 and later in LIDS Student Conference, January
2005.} \\and Causal Transmitter CSI
}
\author{\normalsize  Shashi Borade\hspace{0.1in} Lizhong Zheng\\ \small Laboratory for Information and Decision Systems \\[-5pt]
\small Massachusetts Institute of Technology\\[-5pt]
\small Cambridge, MA 02139, USA\\[-5pt]
\small \{spb,lizhong\}@mit.edu}
\date{}
\maketitle \vspace{1cm}
\begin{abstract}
A wideband fading channel is considered with causal channel state
information (CSI) at the transmitter and no receiver CSI. A simple
orthogonal code with energy detection rule at the receiver (similar
to \cite{telatar}) is shown to achieve the capacity of this channel
in the limit of large bandwidth. This code transmits energy only
when the channel gain is large enough. In this limit, this capacity
without any receiver CSI is the same as the capacity with full
receiver CSI--a phenomenon also true for dirty paper coding. For
Rayleigh fading, this capacity (per unit time) is proportional to
the logarithm of the bandwidth. Our coding scheme is motivated from
the Gel'fand-Pinsker [2,3] coding and dirty paper coding [4].
Nonetheless, for our case, only causal CSI is required at the
transmitter in contrast with dirty-paper coding and Gel'fand-Pinsker
coding, where non-causal CSI is required.

Then we consider a general discrete channel with i.i.d. states. Each
input has an associated cost and a zero cost input ``0'' exists. The
channel state is assumed be to be known at the transmitter in a
causal manner. Capacity per unit cost is found for this channel and
a simple orthogonal code is shown to achieve this capacity. Later, a
novel orthogonal coding scheme is proposed for the case of causal
transmitter CSI and a condition for equivalence of capacity per unit
cost for causal and non-causal transmitter CSI is derived. Finally,
some connections are made to the case of non-causal transmitter CSI
in \cite{pramod}.
\end{abstract}

\section{Introduction}
We consider a wireless fading channel of a large bandwidth $W$. The input $\rvx_{k}[i]$ of band $k$ at time $i$
is related to the output $\rvy_{k}[i]$ as:
\bear\label{eq:channel}\rvy_{k}[i]=\rvh_{k}[i]\rvx_{k}[i]+\rvn_{k}[i]\quad\quad 1\le k\le W,\ \
i\in\{1,2,3\cdots\}\ear where $\rvn_{k}[i]$ is complex circularly symmetric white Gaussian noise of unit
variance. Each $\rvn_{k}[i]$ is independent of all inputs, fading gains, and noise in other bands. The fading
gains $\{\rvh_{k}[i]\}$ are complex Gaussian with variance $1$ and are assumed i.i.d. over time and frequency.
The transmitter has an average power constraint at any time
$i$:\[\sum_{k=1}^W \av{| \rvx_k[i] |^2}{}
\le P\quad\ \ \forall i\]

Note that the channel state at time $i$ is completely described by
the $W$ channel gains $\{\rvh_{k}[i]: 1\le k\le W\}$. We assume (for
reasons discussed later) that at each time $i$, the transmitter
knows this state, i.e. all $W$ fading gains  at that time and the
receiver has no such knowledge. That we are assuming full
transmitter CSI and no receiver CSI.

The case of causal transmitter CSI and no receiver CSI was studied by Shannon for discrete channels
\cite{shannon}. A discrete channel  having $|\cal{X}|$ possible inputs and $|\cal{S}|$ possible states (varying
in i.i.d. manner), can be converted to a discrete memoryless channel of same output alphabet but a larger input
alphabet of size $|\cal{X}|^{|\cal{S}|}$. Capacity of the original channel equals that of this memoryless
channel, which is easier to analyze.

Later, \cite{gp,gamal} studied the following modification of this
scenario. There the channel state for the entire codeword is known
to the transmitter before beginning its transmission. Thus the CSI
is available to the transmitter in a \emph{non-causal} manner,
whereas the receiver has no CSI at all. The optimal code in this
case has a large number of candidate codewords for each message. The
candidate which is suitable to the entire state-sequence spanning
the code-length is used for transmission. More precisely, a
candidate which is jointly typical with the state-sequence is used
for transmission. This motivates our coding scheme for this wideband
fading channel, where the codeword candidate which \emph{benefits}
the most from the state sequence is used for transmission.

For the wideband fading channel above, the capacity without any
receiver and transmitter CSI can be achieved by an orthogonal coding
scheme like Pulse-Position Modulation or Frequency-Shift Keying
\cite{kennedy,telatar}. In the limit of large bandwidth, this
capacity without any CSI equals the capacity with full receiver CSI,
which is $P\log_2 e$  bits per unit time.

For the case of full CSI at both ends, the capacity is achieved by
water-filling which transmits power only when the channel gain is
large enough and this capacity was shown to be essentially $P\log_2
W$ bits per unit time for the Rayleigh fading case \cite{bz}. For
the intermediate case of only transmitter CSI, we wish to combine
these two ideas of orthogonal coding and water-filling. We show that
one can combine these two ideas without loss of optimality, that is,
a code combining these two ideas is shown to achieve the capacity of
this channel. This capacity with only transmitter CSI turns out to
be essentially the same as the capacity ($\approx P\log_2 W$ bits
per unit time) with both transmitter and receiver having CSI. This
is another example where receiver CSI (or lack of it) does not
affect the wideband capacity. In fact, it turns out that this
capacity can be achieved by the proposed code with only one bit of
transmitter CSI for each channel gain without any receiver CSI.

After noting that transmitter CSI can significantly (by a factor of
$\ln W$) increase the capacity of a wideband fading channel
irrespective of receiver CSI, we address the assumption of having
transmitter CSI without any receiver CSI. This may seem to be a
peculiar assumption for a wireless system because the transmitter in
a typical wireless system obtains its CSI through feedback from the
receiver itself. Nonetheless, after feeding back CSI to the
transmitter, the receiver may want to ignore the CSI for multiple
reasons--especially since this does not hurt capacity.

\begin{itemize}\item Ignoring CSI at the receiver may help in
simplifying the decoding algorithm. The structure of the proposed
orthogonal code (for a receiver with no CSI) may simplify the
decoder.
\item Another reason for ignoring receiver CSI comes from the fact that obtaining CSI at the
receiver is intrinsically costly (e.g. in terms of energy spent in
training for CSI). We see later (in section 2) that if receiver CSI
is ignored, obtaining CSI for all channels is not necessary. CSI
needs to be obtained only for a small fraction of channels which
reduces the overall cost of obtaining CSI. This saving in the
channel estimation cost (energy) can bring significant gains in this
wideband system, where the available energy per degree of freedom is
severely limited.\item In addition to less frequent CSI estimation,
ignoring receiver CSI allows for a coarse channel estimation. As
discussed in section 2, the proposed orthogonal code for a receiver
with no CSI requires only one bit of CSI per channel. Obtaining this
single bit of CSI might be easier compared to estimating the exact
channel gain.\end{itemize} The next section describes the coding
scheme and proves its achievable rate.  Section \ref{sec:dmc}
considers a general discrete channel with states. For the case of
causal CSI, an achievable rate for this channel is proved with an
orthogonal code. This is later shown to equal its capacity. In the
last section,  the case of causal transmitter CSI is used to
interpret the case of non-causal transmitter CSI \cite{pramod}.

\section{Capacity achieving scheme}

Our coding scheme is a modification of a scheme like Frequency-Shift Keying scheme or Pulse-Position Modulation,
that is, here the transmitter only transmits if the fading gain is large. The purpose here is to exploit the
channel randomness instead of combating it. We will split the total bandwidth $W$ into $K$ pieces, each of
bandwidth $w=W/K$ and these pieces would be used separately for communication. The available power $P$ is also
equally divided into these pieces. Next, we illustrate our coding scheme for one such piece and analyze its
achievable rate $r$. The total achievable rate would be number of pieces $K$ times the rate per piece $r$. We
will use the notation $f(x)\approx g(x)$ to denote $\lim_{x\rightarrow\infty}\frac{f(x)}{g(x)} =1$.

\begin{figure}[!h]
\centerline{\psfig{figure=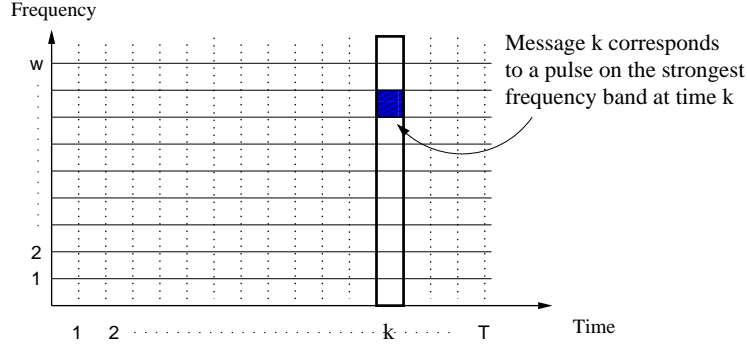,width=10cm,height=4.5cm}}
\caption{Proposed coding scheme: colored symbol indicates energy
transmitted.} \label{fig:fsk}
\end{figure}

The code for such a piece of bandwidth $w$ spans $T$ symbols in time. This code uses each of the $T$ time
indices to denote a message from the set $\{1,2\cdots T\}$. Thus $\ln T$ information nats\footnote{$\ln 2$ nats
$=1$ bit. Hence $\ln T$ nats equals $\log_2 T$ bits. Unless mentioned otherwise, units of rate are nats per unit
time.} are transmitted in time $T$ and hence the code rate is $\ln T/T$ nats per unit time. Say a total of
$\lambda$ energy units are available for this. When message $j$ is to be transmitted, these $\lambda$ energy
units will be transmitted only at time $j$. Moreover, these entire $\lambda$ units of energy are transmitted on
a single frequency band say $f_j$ (see Figure \ref{fig:fsk}). This is the first band where the channel gain for
time $j$ is larger than a threshold $\Phi=\ln w-\ln(2\ln w)$. \bear \label{eq:opttime}f_j=\min\{i:\ |\
\rvh_i[j]\ |^2\ge \Phi\}\ear Note that causal transmitter CSI is enough for this purpose. A type I error is
declared if the channel gains at the time of message $j$ do not cross this threshold for any band.

The decoder calculates the average (over $w$ bands) received energy $E_i$ for each time index $i$
\[E_i=\frac{1}{w}\sum_{k=1}^w |\ \rvy_k[i]\ |^2\ \quad1\le i\le T\]
The time index for which $E_i$ is maximum is declared as the transmitted message. Note that no channel state
information is needed for this decoding method.

Without loss of generality, we will assume that message $1$ was transmitted. A type II error is declared if
$E_i$ the largest for some time other than the time of message $1$. First, we show that probability $P_I$ of
type I error goes to zero for large $w$. Note that type I error occurs if and only if the channel gains of all
$w$ bands at time $1$ are smaller than $\Phi$, that is, the maximum of those $w$ channel gains is smaller than
$\Phi$. Since each $|\ \rvh_k[1]\ |^2$ is exponentially distributed, in the limit of large $w$, their maximum
converges in distribution to \cite{stat}:\bear\ln w+ {\bf z} \quad \textrm{; \ where distribution of } {\bf z}
\textrm{\ is\ \ } P({\bf z}\le Z)= \exp\(-e^{-Z}\)\label{eq:largest}\ear

With our choice of the threshold $\Phi$, probability of type I error is\bear\label{eq:type1}P_I=P\({\bf z}\le -\
\ln(2\ln w)\)=1/w^2\ear Thus probability of type I error vanishes as $w$ tends to infinity. Now we show that
probability $P_{II}$ of type II error also vanishes as $w$ tends to infinity. Assuming that at time $1$, the
channel gain of band $i$ crosses the threshold $\Phi$ (i.e. $f_1=i$), the received symbol in band $i$ at time
$1$ is \bear \rvy_i[1]&=&\rvh_i[1]\sqrt{\lambda}+\rvn_i[1]\\\Rightarrow
 \ \ |\rvy_i[1]|^2&=&|\rvh_i[1]|^2\lambda+|\rvn_i[1]|^2+2\sqrt{\lambda}\Re(\rvh_i[1]\cdot\rvn_i[1]^*)\label{eq:ymag}\\&\ge&\Phi\lambda+|\rvn_i[1]|^2+2\sqrt{\lambda}\Re(\rvh_i[1]\cdot\rvn_i[1]^*)\ear

Using Eq. (\ref{eq:largest}), for large $w$\bear P\(| \rvh_j[1] |^2\ge\ln w +2\ln
w\)&=&1-\exp\(-\frac{1}{w^2}\)\\\label{eq:type3}&\approx &\frac{1}{w^2}\stackrel{w\rightarrow
\infty}{\rightarrow} 0\ear Now assuming $|\rvh_j[1]|^2\le 3\ln w$ yields
\[\Re(\rvh_j[1]\cdot\rvn_j[1]^*)\ge-|\rvh_j[1]||\rvn_j[1]|\ge-\sqrt{3\ln w}|\rvn_j[1]|\]
Note that $|\rvn_j[1]|^2$ is an exponential random variable with mean $1$. Hence \bear P(|\rvn_j[1]|^2\ge2\ln
w)=1/w^2\label{eq:type4}\ear Now assuming $|\rvn_j[1]|^2\le 2\ln w$ implies
\bears\Re(\rvh_j[1]\cdot\rvn_j[1]^*)&\ge&-\sqrt{3\ln w}\sqrt{2\ln w}\\\textrm{Substituting in (\ref{eq:ymag})
implies}\quad\quad\quad\ |\rvy_j[1]|^2&\ge&\Phi \lambda - 2\sqrt{6\lambda}\ln w+|\rvn_i[1]|^2\ears

The above statement may fail if either of the events in Eq. (\ref{eq:type1}),(\ref{eq:type3}),(\ref{eq:type4})
occurs. By union bound, this probability is at most $3/ w^2$. Also note that the noise energy equals the
received energy in all other bands where no energy is transmitted. Hence with at least a probability of
$1-3/w^2$, the average received energy at time $1$ follows:\bear\label{eq:alpha} E_1&\ge& \frac{\Phi \lambda -
2\sqrt{6\lambda}\ln w}{w}+\frac{\sum_{k=1}^w|\rvn_k[1]|^2}{w}\\&\stackrel{\Delta}{=}&
\alpha+\frac{\sum_{k=1}^w|\rvn_k[1]|^2}{w}\ear where $\alpha$ equals the first term in (\ref{eq:alpha}) and is a
non-random variable. By weak law of large numbers, the second term in (\ref{eq:alpha}) converges to $1$ for
large $w$. Average received energy  $E_t$ at any other time $t\ne1$
equals\[E_t=\frac{\sum_{k=1}^w|\rvn_k[t]|^2}{w}\]A type II error occurs when any of these $E_t$ exceeds
$\alpha+1$. Since each $|\rvn_k[t]|^2$ is an exponential random variable with mean $1$, applying Chernoff's
bound on similar lines of \cite{telatar} gives\bear P\(E_t\ge \alpha+1\)&\le&\exp\(-w
L(\alpha)\)\\\textrm{where\quad\quad\quad }\ L(\alpha)&=&
\alpha-\ln(\alpha+1)\label{eq:threshold} \ear Applying union bound
over all wrong messages from $2$ to $T$, we get the following bound
on type II error probability\[P_{II}\le T\exp\(-w
L(\alpha)\)=\exp\(-w(L(\alpha)-\ln T/w)\)\]Since $P_I$ goes to zero
with increasing $w$ as shown before, the overall error probability
vanishes with increasing $w$ if $P_{II}$ also vanishes with
increasing $w$. This happens if \bear \ln T/w &<&
L(\alpha)\\\Rightarrow\quad \ln T/ T &<&\frac{w}{T}
L(\alpha)=\frac{w}{T} \(\alpha-\ln(\alpha+1)\)\label{eq:cond}\ear
Thus the maximum achievable rate\footnote{We have shown that the
error probability of this orthogonal code goes to zero as $w$ goes
to infinity. However, a subtle point is that for showing a rate $\ln
T/T$ is achievable, we have to show that arbitrarily small error
probability can be achieved for a given (but large) $w$ and $T$. As
shown in \cite{bek}, this can be achieved by coding over many blocks
of our orthogonal code by treating the orthogonal code as the inner
code of this concatenated code. The orthogonal code provides an
essentially noiseless discrete memoryless channel (with input
cardinality $T$) for the outer code. Thus a rate $\approx \ln T/T$
can be achieved.} depends on $\alpha$ and hence depends on $\lambda$
(because $\alpha$ equals $\frac{\Phi \lambda - 2\sqrt{6\lambda}\ln
w}{w}$).

Let this scheme be applied only for $\delta$ fraction of the time
where $\delta$ is a suitably chosen parameter. No communication
happens in the remaining fraction of time. Thus if $p$ is the
overall average power available for this piece of bandwidth,
$p/\delta$ is the average power available when communication is
being done. Thus the peakiness denoted by $\delta$ boosts the power
level for actual communication by a factor of $1/\delta$. This boost
in power level is necessary for the success of this orthogonal code.
It ensures that the energy pulse transmitted (for the correct
message) is strong enough to be identifiable at the receiver from
incorrect messages.

Since the time-length of this code is $T$, total transmit energy
$\lambda$ for this code is equal to $Tp/\delta$. Since communication
happens for only $\delta$ fraction of time, the overall maximum
achievable rate is given by
\[r= \delta\frac{w}{T}L(\alpha^*)\quad\quad\textrm{where}\quad\quad\alpha^*=\frac{\Phi p T/\delta -
2\sqrt{6pT /\delta}\ln w}{w}\] Since the total available power  $P$ is divided equally amongst $K$ pieces of the
total bandwidth, power available per piece equals $p=P/K$. We choose $K=\ln w$  and $\delta=\epsi T/w$, where
$\epsi>0$ is a small number. Substituting these values yields $\alpha^*\approx P/\epsi$. Now note that
$L(\alpha)\approx \alpha$ for large $\alpha$. Since $\alpha^*$ can be made arbitrarily large by choosing small
enough $\epsi$, the maximum achievable rate is given by \bears
r&=&\delta\frac{w}{T}L(\alpha^*)\approx\delta\frac{w}{T}\alpha^*\\&\approx&\frac{\epsi T}{w}\frac{w}{T}
\frac{P}{\epsi}\ (=P)\ears Since there are $K=\ln w$ such pieces of bandwidth $w$, the total rate $rK$ equals
$P\ln w$ nats per unit time. The total bandwidth for these $K$ pieces  equals $W=w\ln w $. Noting that $\ln
W\approx \ln w$, the total rate is given by $R\approx P\ln W$ nats per unit time.

This rate expression matches the capacity of this fading channel when the receiver and transmitter both have
full CSI \cite{bz}. This proves that the proposed coding scheme achieves the capacity for this channel with no
receiver CSI. Thus the lack of receiver CSI does not reduce capacity--a phenomenon similar to writing on dirty
paper.

\begin{theorem}
Capacity of the Rayleigh fading wideband channel with causal transmitted CSI and no receiver CSI is achieved by
the proposed coding scheme. In the limit of large bandwidth, this capacity $C\approx P\ln W$ nats per unit time
and is  unchanged if even the receiver has full CSI.
\end{theorem}

Note that as mentioned in Section 1, full transmitter CSI is not
needed for the proposed scheme; only one bit of CSI is enough for
each channel gain $\rvh_i[j]$. This bit indicates whether or not the
channel gain is above the threshold $\Phi$. Also note that CSI is
not needed at every time for this scheme. Since there is no activity
for $(1-\delta)$ fraction of time and only $\delta$ fraction of the
time is used for communication, only this $\delta$ fraction of time
needs CSI and hence the cost of obtaining CSI is significantly
reduced. Since the capacity of a wideband channel with full receiver
and transmitter CSI (at all times) is essentially the same as the
capacity of our channel with only transmitter CSI (for only a
fraction $\delta$ of time), one may want to mimic no receiver CSI
even when it is available!

We can extend above analysis for the case of noisy transmitter CSI,
where the channel gain $\rvh_i[j]$ equals the sum of two independent
complex Gaussian components, $\rvg_i[j]$ and $\rvf_i[j]$, which are
i.i.d. over frequency and time. Transmitter only knows
$\{\rvg_i[j]\}$ and the error $\rvf_i[j]$ is independent of
$\rvg_i[j]$. The variance of the known component is $\beta\in(0,1]$
and hence that of the error is $1-\beta$. A code similar to the
perfect CSI case is employed. For example, if message $1$ is to be
transmitted, the transmitter transmits energy only in the frequency
band where the known channel strength $|\rvg_i[1]|^2$ is larger than
$\beta\Phi$. Thus the threshold for the perfect transmitter CSI case
is reduced by a factor of $\beta$. This scheme can be shown to
achieve a rate of $\beta P\ln W$ nats per unit time. This again
equals the capacity when receiver also has full CSI \cite{bz}. Thus
again receiver CSI is irrelevant for capacity in the limit of large
bandwidth.

\vspace{.1cm} {\bf Remark 1:} Similar results can be proved when distribution of the fading gain $|\rvh_i[j]|^2$
is not exactly exponential but has an exponential tail. If the tail behaves similar to an exponential with mean
$m$, the capacity can be shown to be $mP\ln W$ nats per unit time.

\vspace{.1cm} {\bf Remark 2:} Similar analysis can be performed if
the tail of the fading gain distribution is a polynomial, that is,
$P(|\rvh_i[j]|^2\ge x)\approx x^{-n}$ for some $n>0$. In that case,
the proposed code achieves a rate $R\approx PW^{\frac{1}{n+1}}$ nats
per unit time. This again turns out to be the same as the capacity
when the receiver also has full CSI.

Finally, (on similar lines of \cite{pramod}) we can interpret the proposed scheme in terms of the binning
argument in \cite{gp,gamal}. For the binning interpretation, logarithm of the number of codewords  per message
should equal the mutual information between the state sequence $\{\rvh_j[i]\}$ and the input sequence
$\{\rvx_j[i]\}$. The number of possible codewords per message equals $w$ in our code as energy can be
transmitted on any of the $w$ bands available in a piece. Now note that in our code, the state sequence
completely determines the input sequence for a given message, because we transmit all energy only where the
channel gain first crosses the threshold $\Phi$. Hence the above mutual information equals the input entropy for
a message. Since probability of no frequency band crossing the threshold goes to zero for large $w$ and any of
the $w$ frequency bands are equally likely to cross the threshold, entropy of the input tends to $\log w$. Thus
the binning interpretation is justified as the logarithm of the number of possible codewords per message equals
the mutual information between the input and state sequences.

\section{Capacity per cost with causal transmitter CSI}\label{sec:dmc}
We saw in the previous section how the proposed orthogonal code achieved the capacity of the wide-band fading
channel with no receiver CSI. It also means that the proposed code achieved the capacity per unit cost for that
channel. This section analyzes the case of causal transmitter CSI for a more general channel.

The random variables at time $i\in\{1,2,3\cdots\}$ corresponding to the channel input $X_i$, output $Y_i$ and
channel state $S_i$ take values from the sets ${\cal X} ,{\cal Y}$ and ${\cal S}$ respectively\footnote{Unless
stated otherwise, capital letters denote random variables and small letters denote their values. Notation
$X_1^l$ is used as a shorthand for the sequence $ X_1X_2\cdots X_l$.}. State $S$ defines a channel transition
matrix denoted by $P_{Y|XS}$. The states are assumed to change i.i.d. over time, that is, if  $P_S(\cdot)$
denotes the distribution of $S_i$ then the probability of a state-sequence $s_1^l$ equals
$\prod_{i=1}^lP_S(S_i=s_i)$. Conditioned on the state sequence, the channel is assumed to be memoryless i.e.
\[P(Y_1^l|X_1^l,S_1^l)=\prod_{i=1}^l P_{Y|XS}(y_i|x_i,s_i)\].

Each input $x\in{\cal X}$ incurs a cost $b(x)\in[0,\infty)$. A  zero cost input is assumed to exist and denoted
by $``0"$. In a code of length $l$, the codeword for message $j$ is denoted by the sequence $x_1^l(j)$. A length
$l$ code having $M\in\{1,2\cdots\}$ messages is denoted by a $(l,M,\nu,\epsi)$ code if the average probability
of error is at most $\epsi$ and codeword for every message $j$ satisfies the total cost
constraint\bear\sum_{i=1}^l b(x_i (j))\le \nu\quad\quad0\le j< M\label{eq:cost}\ear The capacity per unit cost
for this channel is defined as in \cite{verdu}.
\begin{definition}
\label{def:cappercost} For a given $0\le \epsi<1$,  rate (in nats) per unit cost $R$ is said to be
$\epsi$-achievable if for all every $\gamma>0$, there exists a $\nu_0$ such that for all $\nu\ge \nu_0$, a
$(l,M,\nu,\epsi)$ code can be found with $\ln M\ge\nu(R-\gamma)$. Rate  per unit cost of $R$ is said to be
achievable if $R$ is $\epsi$-achievable for every $\epsi>0$. Capacity per unit cost is the maximum achievable
rate per unit cost.
\end{definition}
We assume no receiver CSI and causal transmitter CSI, which means that the transmitter gets to know $S_i$ at
time $i$ before transmitting $X_i$. Let $U:\ {\cal S}\rightarrow{\cal X}$ denote a mapping from states to
inputs. This mapping $U$ is equivalent to a vector in ${\cal X}^{|{\cal S}|}$, where its each entry denotes the
input mapped from the corresponding state. Let $P_{Y|U=u}(y)$ denote the output distribution induced when
mapping $U=u$ is chosen, that is, \bear\label{eq:dmc} P_{{Y|U=u}}(y)=\sum_{s\in {\cal
S}}P_S(s)P_{Y|XS}(y|u(s),s)\ear where $u(s)$ denotes mapping of state $s$ under $u$. We next prove the following
theorem.
\begin{theorem}\label{thm:main}Capacity per unit
cost with no receiver CSI and causal transmitter CSI is given by\footnote{We assume that relative entropy and
mutual information are measured in nats i.e. with natural
logarithm.}\[\sup_{u}\frac{D(P_{Y|U=u}||P_{Y|U=0})}{\av{b(X)|U=u}{}}\]where $D(P_{Y|U=u}||P_{Y|U=0})$ denotes
the relative entropy (in nats) between the output distributions induced when mapping $u$ is chosen and when
identically zero mapping is chosen.  $\av{b(X)|U=u}{}$ denotes the average cost incurred when mapping $u$ is
chosen.\[\av{b(X)|U=u}{}=\sum_{s\in {\cal S}}P_S(s)b(u(s))\]
\end{theorem}
\emph{Proof:} We first show an orthogonal coding scheme which achieves the above rate per unit cost. We use the
shorthand $f(n)\doteq g(n)$ to denote $\lim_{n\rightarrow\infty} \frac{\ln f(n)}{\ln g(n)}=1$. Similarly,
$f(n)\stackrel{\cdot}{\le}g(n)$ and $f(n)\stackrel{\cdot}{<}g(n)$ are defined.

Choose a mapping $u:{\cal S}\rightarrow{\cal X}$. Our code of $M$ messages  spans $Mn$ symbols. Each message
corresponds to a non-overlapping interval of length $n$, that is, message $j\in[0,M-1]$ corresponds to
interval\footnote{This means the set of integers from $jn+1$ to $jn+n$.} $[jn+1,jn+n]$. If message $j$ is to be
transmitted, ``$0$" is transmitted at all times except interval $[jn+1,jn+n]$. During each time $i\in
[jn+1,jn+n]$, input $u(S_i)$ is transmitted. This requires only causal CSI at the encoder.

Assuming message $j$ was transmitted, the output distribution at each time in interval $[jn+1,jn+n]$ is given by
$P_{Y|U=u}$. Outputs in all other intervals are distributed as $P_{Y|U=0}$. For each of the $M$ intervals of
length $n$, the decoder finds the empirical output distribution of that interval. Let $P_Y^k$ denote this
empirical distribution for interval $[kn+1,kn+n]$. The interval $k$ for which $D(P_Y^k||P_{Y|U=0})$ is larger
than a threshold\footnote{This threshold is  finite if the support of $P_{Y|U=u}$ is contained in that of
$P_{Y|U=0}$. This threshold is not finite if there exists an output (say $\hat{y}$) which can only occur with a
non-zero input. The decoding in that case would be  easy because only the correct interval can have the output
$\hat{y}$.} $\Phi=D(P_{Y|U=u}||P_{Y|U=0})-\delta$ is declared as the transmitted message, where $\delta$ is a
chosen small number. An error is declared when none or multiple such intervals exist.

First kind of error occurs if the divergence $D(P_Y^j||P_{Y|U=0})$ for the correct interval does not exceed the
threshold $D(P_{Y|U=u}||P_{Y|U=0})-\delta$. By Sanov's theorem (e.g. \cite{sanov}), this probability goes to
zero exponentially fast in $n$. Hence this probability of error of first kind is smaller than $\epsi/3$ for all
$n\ge n_1$ for some $n_1>0$.

The second kind of error occurs if the divergence $D(P_Y^k||P_{Y|U=0})$ for a wrong interval $k\ne j$ exceeds
the threshold. Again applying Sanov's theorem implies \bear\label{eq:exp}P\(D(P_Y^k||P_{Y|U=0})>\Phi\)\doteq
\exp(-n\Phi)\ear By union bound, the probability $P_{II}$ that any of the $M-1$ wrong intervals crosses this
threshold is bounded by\[P_{II}\stackrel{\cdot}{\le} M\exp(-n\Phi)\]If we choose $M=\exp(n(\Phi-\delta))$,
probability $P_{II}$ also goes to zero exponentially as $\exp(-n\delta)$. Thus the probability of error of
second kind is smaller than $\epsi/3$ for all $n\ge n_2$ for some $n_2>0$.

By i.i.d. nature of the states and the law of large numbers, total cost for each message is smaller than
$n(\av{b(X)|U=u}{}+\delta)$ with (at least) a probability of $1-\epsi/3$ if $n\ge n_3$ is chosen for some
$n_3>0$.

Thus even if an error of third kind is declared if the total cost of the codeword exceeds the threshold, the
total probability of any kind of error is less than $3(\epsi/3)$ for all $n\ge\max(n_1,n_2,n_3)$. Thus for any
$\epsi>0$ and $\gamma>0$, we can choose small enough $\delta$ such that \bears\ln
M=n(\Phi-\delta)\\&=&n\(D(P_{Y|U=u}||P_{Y|U=0})-2\delta\)\\&>&n(\av{b(X)|U=u}{}+\delta)\(\frac{D(P_{Y|U=u}||P_{Y|U=0})}{\av{b(X)|U=u}{}}-\gamma\)\ears
and the probability of error is smaller than $\epsi$ for $n\ge\max(n_1,n_2,n_3)\stackrel{\Delta}{=}n_*$.
Substituting $\nu=n(\av{b(X)|U=u}{}+\delta)$ and $\nu_0=n_*(\av{b(X)|U=u}{}+\delta)$ in the definition of the
rate per unit cost proves that the proposed orthogonal code achieves a rate per unit cost of
$D(P_{Y|U=u}||P_{Y|U=0})/\av{b(X)|U=u}{}$.

{\bf Remark 3}: Note the similarity of this scheme with the coding scheme in previous section for the wideband
fading channel. In particular, note that the probability of an incorrect interval crossing the threshold $\Phi$
is given by $\exp(-n\Phi)$, similar to (\ref{eq:threshold}). For these reasons, one can interpret the divergence
$D(P_Y^i||P_{Y|U=0})$ for  interval $i$ as the discrete channel analogue of the average received energy $E_i$
for the wideband fading channel.

\emph{ Proof of converse}: We first note the following upper bound in \cite{verdu} on capacity per unit cost of
a discrete memoryless channel with input $V$ and output $Z$\bear\label{eq:capcost}\sup_v
\frac{D(P_{Z|V=v}||P_{Z|V=0})}{c(v)}\ear where $P_{Z|V=v}$ denotes the output transition probability for input
$v$, $c(v)$ denotes the cost of input $v$ and $V=0$ denotes the zero cost input.

Now recall Shannon's idea \cite{shannon} that this channel with causal transmitter CSI and i.i.d. states can be
thought as a discrete memoryless channel (DMC) with the same output alphabet but a larger input alphabet. The
input alphabet $U$ of that equivalent DMC corresponds to a mapping from ${\cal S}$ to ${\cal X}$ and thus its
cardinality equals $|{\cal X}|^{|{\cal S}|}$. An input $U$ of this DMC is equivalent to a vector in ${\cal
X}^{|{\cal S}|}$ made up of contingent inputs (from ${\cal X}$) for each state $s\in {\cal S}$.  A code for the
DMC can be converted to a code for causal transmitter CSI channel as follows. If the symbol $u_i$ was
transmitted at time $i$ on the DMC, the transmitter with causal CSI transmits input $u_i(S_i)$ at time $i$ after
observing state $S_i$.

This DMC is a cascade of two memoryless parts. First part chooses the state $S_i$ with distribution $P_S$ and
picks the corresponding contingent input $u_i(S_i)\in{\cal X}$ from the transmitter. Second part is similar to
our original channel of interest, which emits the output based on the state $S_i$ and the input $u_i(S_i)$
according to the distribution $P_{Y|XS}(\cdot|u_i(S_i),S_i)$. The output distribution of this DMC conditioned on
the input $u$ is given by $P_{Y|U=u}$ in (\ref{eq:dmc}).

Finally, note that $\av{b(X)|U=u}{}$ denotes the (average) cost incurred due to choosing the DMC input $u$. The
converse follows by applying (\ref{eq:capcost}) after replacing $P_{Z|V=v}$ by $P_{Y|U=u}$, $P_{Z|V=0}$ by
$P_{Y|U=0}$ and $c(v)$ by $\av{b(X)|U=u}{}$. A more detailed converse is proved in the appendix.

We could also prove the direct part of this theorem using above method of conversion to a DMC. However, the
earlier detailed proof is expected to be more insightful in view of writing on fading paper.
\section{Discussion}
For the wideband fading channel, we noted that the capacity with
causal CSIT\footnote{CSIT: Acronym for transmitter CSI.} was the
same as that with non-causal CSIT. Equivalently, the capacity per
unit cost was the same with causal or non-causal CSIT. Similar
phenomenon can be shown for the AWGN channel with additive Gaussian
interference known at the transmitter by a modification of the
scheme in \cite{pramod}. We want to understand whether these are
isolated examples (of the equivalence of capacity per unit cost with
causal and non-causal CSIT) or they are special cases of a general
class.

If with causal or non-causal CSIT, the capacities (for any given cost constraint) are the same for a channel;
then it is easy to show that the capacities per unit cost would also be the same for that channel. This is
because for a channel with a $0$ cost alphabet, capacity per unit cost is given by the slope of the capacity vs.
cost curve at $0$.

More interesting problem is to characterize the class of channels for which the capacity per unit cost is the
same with causal or non-causal CSIT, but the capacity vs. cost curves are not the same for causal and non-causal
CSIT. Above mentioned wideband AWGN channel and wideband fading channel with additive interference are two such
channels.
\subsection{Review of the non-causal transmitter CSI case}
We briefly summarize the coding scheme that achieves the capacity per unit cost with non-causal CSIT
\cite{pramod}. This code of $M$ messages spans $Mqn$ symbols. Each message in this orthogonal code corresponds
to a separate interval of length $qn$. For transmitting a message $j$, non-zero symbols can be only transmitted
in the $j$'th interval of length $qn$. This message interval of length $qn$ can be thought as the set of $q$
subintervals, each of length $n$.

A distribution of states $\hat{P}_S(\cdot)$ is chosen beforehand. Out of these $q$ subintervals in the interval
for message $j$, the subinterval whose empirical distribution is like $\hat{P}_S(\cdot)$ is chosen. More
precisely, the divergence of the empirical distribution of this subinterval with respect to $\hat{P}_S(\cdot)$
should be small enough. Since the actual distribution of states is $P_S$, the probability of a subinterval
having distribution like $\hat{P}_S(\cdot)$ is essentially (in $\doteq$ sense) given by
$\exp\(-nD(\hat{P}_S||{P}_S)\)$. We can find such a subinterval with high probability if the number of
subintervals per message interval is \bear\label{eq:fraction}q\doteq\exp\(nD(\hat{P}_S||{P}_S)\)\ear Non-zero
symbols are only transmitted in this subinterval. A mapping
 $u:{\cal S}\rightarrow{\cal X}$ is also chosen beforehand. Similar to previous section, input $u(s)$ is
 transmitted for state
 $s$ in this subinterval. The output distribution in this subinterval would be
 \bear \label{eq:yhat}\hat{P}_Y(y)=\sum_{s\in{\cal S}}\hat{P}_S(s)P_{Y|XS}\(y|u(s),s\) \ear
Output distribution in all other subintervals (where only input $0$ is transmitted) is $P_{Y|U=0}$
\[P_{Y|U=0}(y)=\sum_{s\in{\cal S}}{P}_S(s)P_{Y|XS}\(y|0,s\)\]

Note that non-zero symbols are transmitted in a small fraction ($1/q$) of the interval corresponding to message
$j$. Note from (\ref{eq:fraction}) that this fraction decays exponentially to $0$ with increasing $n$. Also note
that non-causal CSIT is necessary to determine the subinterval having empirical distribution like
$\hat{P}_S(\cdot)$.

At the receiver, empirical distribution is found for all the $q$ subintervals for each of the $M$ message
intervals. If one of these $Mq$ subintervals has distribution like $\hat{P}_Y$ in (\ref{eq:yhat}), the message
interval containing that subinterval is declared as the transmitted message. An error is declared otherwise.
Since every wrong  subinterval is distributed as $P_{Y|U=0}$, probability of its having an empirical
distribution as $\hat{P}_Y(y)$ is essentially $\exp(-nD(\hat{P}_Y||P_{Y|U=0}))$. Thus by union bound, the
probability of a wrong subinterval having output distribution $\hat{P}_Y(y)$
is\[Mq\exp(-nD(\hat{P}_Y||P_{Y|U=0}))=M\exp\(nD(\hat{P}_S||{P}_S)-nD(\hat{P}_Y||P_{Y|U=0})\)\] Choosing
$M\doteq\exp\(n(D(\hat{P}_Y||P_{Y|U=0})-D(\hat{P}_S||{P}_S))\)$ can ensure that probability error vanishes with
large $n$. By law of large numbers, the total cost incurred for transmission is  essentially
$n\av{b(u(S))}{\hat{P}_S}$ where\[\av{b(u(S))}{\hat{P}_S}=\sum_{s\in{\cal S}}\hat{P}_S(s)b(u(s))\] Thus the rate
per unit cost achieved by this scheme is\bear\label{eq:capnc}\frac{\ln
M}{n\av{b(u(S))}{\hat{P}_S}}\doteq\frac{D(\hat{P}_Y||P_{Y|U=0})-D(\hat{P}_S||{P}_S)}{\av{b(u(S))}{\hat{P}_S}}\ear
Optimizing above expression over the choice of $\hat{P}_S$ and $u(\cdot)$ can be shown to yield the capacity per
unit cost for this channel with non-causal CSIT. We denote an optimum choice by  $\hat{P}_S^*$ and $u^*(\cdot)$,
respectively.
\subsection{Adapting to the causal transmitter CSI case}
With causal CSIT, transmitter does not a priori know the subinterval having empirical state distribution
$\hat{P}_S$. To overcome this issue, let there be only one subinterval per interval i.e. let $q=1$. Thus each
message corresponds to an interval of length $n$. Now a fraction $\theta$ is chosen by the transmitter. The
transmitter can only transmit energy (non-zero symbols) in a fraction $\theta$ of message interval. For each
state $s\in {\cal S}$, the transmitter will transmit input $u(s)$ for the first $n\theta\hat{P}_S(s)$
occurrences of state $s$. Thus the states where energy is transmitted will have an empirical distribution
$\hat{P}_S$. Since the actual state distribution is $P_S$,  (by law of large numbers) an interval of length $n$
will have $n\theta\hat{P}_S(s)$ occurrences of state $s$ only if \bear\label{eq:statecond}\theta\hat{P}_S(s)\le
P_S(s) \quad \forall s\in{\cal S}\quad\Rightarrow \theta\le\inf_{s\in{\cal S}}\ \{{P}_S(s)/\hat{P}_S(s)\}\ear
Note that the $n\theta$ symbols where energy is transmitted need not be in a contiguous block. Again by law of
large numbers, the total cost incurred in this procedure is (in $\approx$ sense) essentially
$n\theta\av{b(u(S))}{\hat{P}_S}$.

At the decoder, for each message interval of length $n$, the empirical output distribution is found for all
${n\choose n\theta}$ subsequences\footnote{The term subinterval is reserved for contiguous blocks of symbols,
whereas a subsequence need not be contiguous.} of length $n\theta$. Out of these $M{n\choose n\theta}$
subsequences, if all the subsequences having distribution like $\hat{P}_Y$ in (\ref{eq:yhat}) belong to a single
message interval, the message corresponding to that message interval is declared as the transmitted message. An
error is declared if more than one or none of the message intervals have such subsequences. By law of large
numbers, the correct subsequence of length $n\theta$ where energy is transmitted will have an empirical output
distribution like $\hat{P}_Y$ with high probability for large $n$. A length $n\theta $ subsequence in an
incorrect message interval will have an empirical output distribution like $\hat{P}_Y$ with probability $p_1$
given by
\[p_1\doteq\exp\(-\theta nD(\hat{P}_Y||P_{Y|U=0})\)\]Applying union bound, the probability of a subsequence of
an incorrect message interval having empirical output distribution like $\hat{P}_Y$ is bounded by $M{n\choose
n\theta}p_1$. Hence vanishing error probability can be achieved if \bears M\doteq\frac{1}{{n\choose
n\theta}p_1}&=&\frac{\exp\(\theta nD(\hat{P}_Y||P_{Y|U=0})\)}{{n\choose n\theta}}\\&\doteq&\frac{\exp\(\theta
nD(\hat{P}_Y||P_{Y|U=0})\)}{\exp\(nH_b(\theta)\)}\quad\quad\textrm{by Sterling Approximation}\\&=&\exp\(
n(\theta D(\hat{P}_Y||P_{Y|U=0})-H_b(\theta))\)\ears

The rate per unit cost achieved by this scheme equals\bears\frac{\ln
M}{n\theta\av{b(u(S))}{\hat{P}_S}}&=&\frac{\theta
D(\hat{P}_Y||P_{Y|U=0})-H_b(\theta)}{\theta\av{b(u(S))}{\hat{P}_S}}\\&=&\frac{
D(\hat{P}_Y||P_{Y|U=0})-H_b(\theta)/\theta}{\av{b(u(S))}{\hat{P}_S}}\ears
For this to equal the capacity per unit cost with non-causal CSIT in
(\ref{eq:capnc}), an optimum $\hat{P}^*_S$ maximizing
(\ref{eq:capnc}) should
satisfy\[H_b(\theta)/\theta=D(\hat{P}^*_S||{P}_S)\]Now note that
(\ref{eq:statecond}) implies\bears
D(\hat{P}^*_S||{P}_S)&=&\sum_{s\in{\cal
S}}\hat{P}^*_S(s)\ln\(\frac{\hat{P}^*_S}{{P}_S}\)\\&\le&\sum_{s\in{\cal
S}}\hat{P}^*_S(s)\ln\frac{1}{\theta}=\ln(1/\theta)\\&\le&H_b(\theta)/\theta\ears
Last step is met with equality either when $\theta=1$ or when
$\theta$ tends to zero. For equality in second step, we need
${\hat{P}^*_S}/{{P}_S}=1/\theta$ for all states having
$\hat{P}_S^*(s)>0$.

Case when $\theta=1$ corresponds to $\hat{P}^*_S={P}_S$. By law of large numbers, empirical distribution for
each interval would be $P_S$ with high probability. Thus the non-causal nature of transmitter CSI is rendered
useless in this case because only $1$ subinterval (i.e. $q=1$)  suffices per message interval.

Thus this coding scheme gives the following sufficient condition for
the capacity per unit cost with causal or non-causal CSIT to be the
same.

\begin{theorem}\label{lemma}Let $\mu$ denote $\inf_{s\in{\cal S}}\{\frac{{P}_S(s)}{\hat{P}^*_S(s)}\}$ for an
optimum $\hat{P}^*_S$ that achieves the capacity per unit cost for
non-causal CSIT in (\ref{eq:capnc}). For equivalence of capacity per
unit cost with causal and non-causal CSIT, $\mu$ should either be
arbitrarily small or be equal to $1$. Moreover, for all states in
support of $\hat{P}_S$ (i.e. states having $\hat{P}_S(s)>0$) should
achieve\footnote{If $\mu$ tends to $0$, this clause can be relaxed
as long as $H_b(\mu)/\mu$ approaches $D(\hat{P}^*_S||{P}_S)$.} the
infimum $\mu={P}_S(s)/\hat{P}_S(s)$.\end{theorem} If $\mu$ tends to
zero, the divergence $D(\hat{P}^*_S||{P}_S)$ should tend to infinity
to satisfy the above condition. In other words, its arbitrarily rare
to observe the source distribution where energy is transmitted. This
is because (by Sanov's theorem) the larger $D(\hat{P}^*_S||{P}_S)$
is, the rarer it is to have empirical distribution like
$\hat{P}^*_S$ when actual state distribution is ${P}_S$.

Note that the wideband fading channel and the wideband writing on
dirty paper \cite{pramod} satisfy the above Lemma, which gaurantees
that capacity per unit cost is the same with causal or non-causal
CSIT. The fraction of states $\theta$ where energy was transmitted
was arbitrarily small there. Thus the above Lemma explains some
reasons for the equivalence of the capacity per unit cost with
causal and non-causal CSI for those channels.

With this background,  we  revisit the capacity achieving scheme for the non-causal CSIT case. The state vector
of each length $n$ subinterval  can be viewed as a superstate of cardinality $|{\cal S}|^n$. Now each message in
the code corresponds to an interval consisting of $q$ super-symbols (or subintervals).

A subinterval of empirical distribution $\hat{P}_S$ corresponds to a superstate with probability
$\exp(-nD(\hat{P}_S||P_S))$. Energy is only transmitted in these rare subintervals.  Non-causal CSIT of a length
$n$ subinterval in the original channel corresponds to causal CSIT in the super-channel. The idea of
subintervals has thus converted the channel with non-causal CSIT to a channel with causal CSIT.

The causal CSIT channel (with superstates) has some arbitrarily rare superstates where energy is transmitted for
achieving capacity per unit cost. Hence by Lemma \ref{lemma}, the capacity per unit cost for this super-channel
is the same with causal or non-causal CSIT. Since non-causal CSIT for the super-channel also means non-causal
CSIT for the original channel, the capacity per unit cost of the super-channel for causal CSIT equals the
capacity per unit cost of the original channel for non-causal CSIT. Thus even if  Lemma \ref{lemma} is not
satisfied for the original channel directly, the idea of subintervals achieves the non-causal capacity per unit
cost by converting the original channel to a super-channel for which Lemma \ref{lemma} is satisfied. This is
achieved by providing arbitrarily rare (super)states for transmitting energy.

\section*{Acknowledgements} Thanks to Robert Gallager for suggesting a simple On-Off fading channel,
which prompted the writing on fading paper scheme. Shashi Borade
also acknowledges numerous insightful comments and suggestions by
Ashish Khisti.

\section*{Appendix: Proof of Converse of Theorem \ref{thm:main}}

We use a technique similar  to \cite{verdu}\cite{pramod}, which adapts a converse for capacity to a converse for
capacity per unit cost. For a code of length $l$, by Fano's inequality we know the following necessary condition
for transmitting a message $m$ chosen uniformly out of $M$ possible messages with error probability smaller than
$\epsi$.
\[ (1-\epsi)\ln M\le I(m;Y_1^l)+H_b(\epsi)\]
where $H_b(\cdot)$ denotes the entropy of a binary variable as a function of its probability of being $1$. From
the converse for the capacity of a channel with causal transmitter CSI, we know that
\cite{shannon}\bear\label{eq:nece} I(m;Y_1^l)\le \sum_{i=1}^l I(U_i;Y_i) \ear where $I(U_i;Y_i)$ denotes the
mutual information between the state to input mapping  $U_i$ and output $Y_i$. The mapping $U_i$ is considered
as a random variable of cardinality $|{\cal X}|^{|{\cal S}|}$. The mutual information can be thought as the
mutual information of a channel with input $U_i$ and output $Y_i$, where the output transition probability for
input $u$ is given by $P_{Y|U=u}$ in (\ref{eq:dmc}).

Now we introduce a time-sharing random variable $Q$, which is independent of all other variables and is
uniformly distributed over integers from $1$ to $l$. This gives the following upper bound\bears \sum_{i=1}^l
I(U_i;Y_i)&=&lI(U_Q;Y_Q|Q)\\&=&l\(I(U_Q,Q;Y_Q)-I(Q;Y_Q) \)\\&\le& l I(U_Q,Q;Y_Q)\ears Defining ${U}=(U_Q,Q)$ and
defining $Y=Y_Q$, we get the upper bound on $(1-\epsi)\ln M$ as $lI({U};Y)+H_b(\epsi)$.

Now assume a weaker average cost constraint instead of the per codeword cost constraint in (\ref{eq:cost}) as
follows
\[\av{\sum_{i=1}^l b(X_i)}{}\le \nu \]Using the time sharing variable and later replacing $X_Q$ by $X$ gives
\[\nu\ge\av{\sum_{i=1}^l b(X_i)}{}= \sum_{i=1}^l\av{b(X_Q)|\ Q=i}{}=l\av{b(X_Q)}{}=l\av{b(X)}{}\]Combining this
with the upper bound on $(1-\epsi)\ln M$ gives\[\frac{\ln M}{\nu} \le \frac{I({U};Y)+H_b(\epsi
)/l}{(1-\epsi)\av{b(X)}{}}\]As $\epsi$ can be arbitrarily small and $l$ can be arbitrarily large, we get $\ln
M/\nu\le I(U;Y)/\av{b(X)}{}$ as the necessary condition for arbitrarily small error probability on a code. Thus
a code with arbitrarily small error probability $\epsi$ must satisfy $\frac{\ln M}{\nu}\le\frac{
I(U;Y)}{\av{b(X)}{}}$ for some choice of random variable $U$ (which denotes a mapping from states to inputs).

Now note that mutual information $I(U;Y)$ can be written as \bear I(U;Y)&=&\sum_u
P_U(u)D(P_{Y|U=u}||P_Y)\\&=&\sum_u P_U(u)D(P_{Y|U=u}||P_{Y|U=0})-D(P_{Y|U=0}||P_Y)\\&\le&\sum_u
P_U(u)D(P_{Y|U=u}||P_{Y|U=0}) \label{eq:ineq}\ear where $P_{Y|U=0}$ indicates the output distribution when the
state to input mapping is identically zero i.e. when $0$ input is transmitted for any state. Also note that the
expected cost can be written as\bears \av{b(X)}{}&=&\sum_u P_U(u) \av{b(X)|U=u}{}\ears Combining this with
(\ref{eq:ineq}) gives that any code with arbitrarily small error probability should satisfy \[\frac{\ln
M}{\nu}\le\frac{\sum_u P_U(u)D(P_{Y|U=u}||P_{Y|U=0})}{\sum_u P_U(u) \av{b(X)|U=u}{}}\]for some distribution
$P_U(\cdot)$ of the state to input mapping $U$. In other words,\bears\frac{\ln
M}{\nu}&\le&\sup_{P_U}\frac{\sum_u P_U(u)D(P_{Y|U=u}||P_{Y|U=0})}{\sum_u P_U(u)
\av{b(X)|U=u}{}}\\&=&\sup_u\frac{D(P_{Y|U=u}||P_{Y|U=0})}{\av{b(X)|U=u}{}}\quad\quad\quad\textrm{Q.E.D.}\ears

\end{document}

%% file: preamble.tex
\renewcommand{\(}{\left(}
\renewcommand{\)}{\right)}

\newcommand{\rvg}{{\bf g}}

\newcommand{\rvf}{{\bf f}}
\newcommand{\av}[2]{{{{\cal E}_{{#2}}\left[{#1}\right]}}}

\newcommand{\blist}{    \begin{list}{$\bullet$}{\topsep 0.0in \partopsep
0.0in
                \itemsep 0.05in \parsep
                0.0in \leftmargin 0.3in}}
\newcommand{\elist}{\end{list}}

\newcommand{\beq}{\begin{equation}}
\newcommand{\be}{\begin{equation}}
\newcommand{\eeq}{\end{equation}}

\newcommand{\bear}{\begin{eqnarray}}
\newcommand{\ear}{\end{eqnarray}}
\newcommand{\bears}{\begin{eqnarray*}}
\newcommand{\ears}{\end{eqnarray*}}

\newcommand{\epsi}{{\epsilon}}

\newcommand{\rvy}{{\bf y}}
\newcommand{\rvx}{{\bf x}}

\newcommand{\rvh}{{\bf h}}

\newcommand{\rvn}{{\bf n}}

\newtheorem{theorem}            {Theorem}

\newtheorem{definition}         [theorem]{Definition}


%% file: fading-paper.bbl
\begin{thebibliography}{99}
\bibitem{shannon}
C. E. Shannon, ``Channels With Side Information at the Transmitter,'' \emph{IBM Journal Research and
Development}, Vol. 2, pp. 289--293, 1958.
\bibitem{gp}
S. Gel'fand and M. Pinsker, ``Coding for Channel with Random Parameters,'' \emph{Problems of Control and
Information Theory}, Vol. 9, No. 1, pp. 19---31, 1980.
\bibitem{gamal}
C. Heegard and A. Gamal, ``On the Capacity Computer Memory with Defects,'' \emph{IEEE Transactions on
Information Theory}, Vol. 29, No. 5, pp. 731--739, September 1983.
\bibitem{costa}
M. Costa, ``Writing on dirty paper,'' \emph{IEEE Transactions on Information Theory}, vol. 29, no. 3, pp.
439-441, May 1983.
\bibitem{kennedy}
R. Kennedy,\emph{ Fading Dispersive Communication Channels}, New York: Wiley-Interscience, 1969,
\bibitem{telatar}
I.~E.~Telatar and D. N. C. Tse, ``Capacity and mutual information of wideband multipath fading channels,'' {\em
IEEE Trans. on Information Theory}, vol.~46, pp.~1384--1400, July 2000.
\bibitem{bz}
S. Borade and L. Zheng, ``Wideband fading channels with feedback,'' \emph{Proc. Allerton Conference}, Sept.
2004.
\bibitem{pramod}
T. Liu and P. Vishwanath, ``Opportunistic orthogonal writing on dirty paper,'' submitted to {\em IEEE Trans. on
Information Theory}.
\bibitem{verdu}
S. Verdu, ``On channel capacity per unit cost,'' \emph{IEEE Transactions on Information Theory}, Vol. 36, No. 5,
pp. 1019--1030, September 1990.
\bibitem{stat}
H. David, \emph{Order Statistics}, 1st ed., New York: Wiley, 1970.
\bibitem{sanov}
P. Dupuis, R. Ellis, \emph{A Weak Convergence Approach to the Theory of Large Deviations}, Wiley-Interscience,
1997.
\bibitem{bek}
 ``Writing on dirty paper with causal CSI and effects of intereference tail''\emph{In preparation}.


\end{thebibliography}
